# Breaking Political Filter Bubbles via Social Comparison

\* **Motahhare Eslami**
University of Illinois at Urbana-Champaign
Champaign, IL 61820, USA
eslamim2@illinois.edu

\* **Nouran Soliman**
Arab Academy for Science, Technology and Maritime Transport
Alexandria, Egypt
nouranm.soliman@student.aast.edu

**Karrie Karahalios**
University of Illinois at Urbana-Champaign
Champaign, IL 61820, USA
kkarahal@illinois.edu

## Abstract
Online social platforms allow users to filter out content they do not like. According to selective exposure theory, people tend to view content they agree with more to get more self-assurance. This causes people to live in ideological filter bubbles. We report on a user study that encourages users to break the political filter bubble of their Twitter feed by reading more diverse viewpoints through social comparison. The user study is conducted using political-bias-analyzing and Twitter-mirroring tools to compare the political slant of what a user reads and what other Twitter users read about a topic, and in general. The results show that social comparison can have a great impact on users' reading behavior by motivating them to read viewpoints from the opposing political party.

## ACM Classification Keywords
H.5.m [Information interfaces and presentation (e.g., HCI)]: Miscellaneous

## Introduction
Previous research highlights that people usually avoid cognitive dissonance as they prefer to access agreeable in-formation [4, 10, 18, 1, 9, 5]. As proposed by the selective exposure theory, people prefer to view information that self-assures them about their information by supporting their beliefs without the need to reevaluate their stances [7]. Polarization can be caused by interacting with people with similar stances [20]. Selective exposure produces a similar impact which causes audiences to be more extreme and to struggle to find a common point.

In addition to selective exposure, as a lot of political news sources are becoming available, many social and news feeds employ filtering algorithms to prevent information overload. In doing so, these algorithms usually learn to filter out the content that users do not like or do not engage with enough. This reinforces the impacts of selective exposure, encapsulates users in ideological "filter bubbles" that isolate users from diverse viewpoints [17] and causes less creative thinking [19]. As a result, skills like problem solving and decision making by groups or individuals can deteriorate [14, 13]. This raises the challenge of encouraging users to break their online political filter bubble and read more diverse political news.

To address this challenge, several methods have been introduced [16, 2, 15, 12, 11] that nudge users to read different political view points. This nudging practice usually is done by self-comparison, i.e., showing users what they read and what they could read. These studies, however, are shown to be only somewhat effective due users' strong preference for reading primarily agreeable news.

In this study, we employ *social comparison* (rather than self-comparison) as a nudging technique to motivate users to break their political filter bubble. We designed a tool to replicate a user's Twitter feed enhanced with visualizations that compare the political slants of what the user reads and what other Twitter users read. We found that exposing users to what other Twitter users read encouraged them to read more diverse political opinions, particularly when a user found that what other users read about a political topic had an opposing political slant from what she read. These results are promising to further study the impacts of social comparison in encouraging users to read more diverse news.

## Related Work
The research studies related to breaking the filter bubble mainly employed one of these methodologies: 1) self-comparison of reading activity or 2) interaction with other users. We discuss these methodologies below and then describe the theory of social comparison upon which we build our methodology.

### *Self-Comparison of Reading Activity*
The main line of research in breaking the filter bubble revolves around nudging users to read more diverse viewpoints via comparing what they read and what they could read. For example, NewsCube [15] classifies news articles' political slants so the user can compare articles via different political perspectives to each other. Although, NewsCube's evaluation suggested that readers formed more balanced opinions, these results might not transfer to other cultures as the experiment was done only on Korean news. In another study, Munson et al. [11] developed a browser extension that helped users monitor their political leanings of weekly and all-time readings. However, while the tool helped a few users improve their reading balance, it did not affect most of them. To understand the reasons behind this, Munson and Resnick [12] performed another experiment in which they found that users are usually either challenge-averse or diversity-seeking; those that are challenge-averse will not be affected by such nudging practices.

### Interaction with Other Users

In some studies, researchers went beyond self-comparison, and tried to engage users with opposite political perspectives with each other. For example, the Electronic Dialogue Project [16] created monthly interactions between users from various backgrounds to discuss political topics. The study found that the discussions had positive effects on engagement of participants together. In another study, the Political Blend application [2] encouraged users with opposite political slants to meet over coffee. To incentivize meeting, a coupon was created when both appeared at the cafe. The study reflected some evidence that social interactions in neutral contexts can help with exchanging diverse political ideas constructively.

### Social Comparison Theory

In this paper, we try to address the challenges in the previous studies on breaking the filter bubble via employing social comparison theory. The theory indicates that each individual has a drive to accurately self-evaluate herself [3] via comparing herself to others to define the self. Social comparison is considered a way of self-enhancement [6, 21]. We use this theory to understand if social comparison is effective to encourage users to break their filter bubble.

## Study Design

In this study, we designed a tool, *BubbleBreaker*, which compares the political slant of what a Twitter user reads with what other Twitter users read about a political topic. We then conducted an interview with nine Twitter users to engage them with this tool and compared the political bias of their Twitter feed versus other users.

### BubbleBreaker

We first collected the top 200 tweets from each participant's home feed using the Twitter API. We then manually determined the topic of a tweet, and labelled it as political or apolitical. Next, we calculated two biases for each political tweet in the user's feed: a) the political slant of a tweet itself, and b) the political slant of what other users read about that tweet's topic on Twitter (Twitter world bias). In addition, we computed an overall feed political bias for each user. We explain each of these biases below.

*A Tweet's Political Bias:* We employed the method used by kulshrestha et. al. [8] to calculate the political bias of an individual tweet. This method estimates the political bias of a tweet's author by matching her topical interests to the interests of generated democratic and republican representative sets. This method highlights that the match between the source bias (bias of author of tweet) and the content bias (bias of tweet text) of the tweet can reach 75% or more.

To display this bias, we added a bar with a gradient coloring beside each political tweet that shows the possible political bias range of a tweet. We then added a pointer in the shape of the author of the Tweet to display what the tweet's bias is (Figure 1).

*Twitter World Bias:* We define the Twitter world bias on a certain topic to be the overall bias of the top 20 search results that Twitter search returns on that topic. The reason for this choice was that if a user wanted to read about a topic on Twitter, Twitter search is the main way to retrieve information about that topic. To calculate this bias, we first used the topic title previously given to each political tweet to retrieve the top 20 search results on that topic via the Twitter API. We then computed the political bias of each tweet in the search results, via the method described in the previous section. Finally, we calculated an overall bias from these biases calculated for search results. This overall bias is computed via the Average Precision metric which takes into account the ranking of a tweet in the search results as in [8]. In this method, the higher the rank of a tweet is in the search results, the more weight its bias is given in calculating the overall bias of the list.

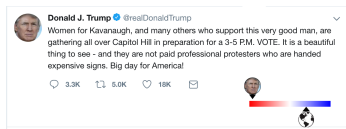

**Figure 1:** Political bias is defined based on a 7 point Likert scale with values from 1 to 7; 1 corresponds to strong republican, 4 is the neutral point and 7 is strong democrat. This scale is represented as a bar with gradient coloring from red to white then from white to blue. The intensity of the color is used to reflect the value of the bias. The bias of the tweet is marked on the bar by a pointer attached to the profile picture of the author of the tweet and the world's bias is marked on the bar by another pointer attached to the picture of a globe to reflect the concept of social comparison with the Twitter world.

To represent the world bias, we added a world icon to the gradient bar adjacent to each political tweet. The Twitter world globe icon is also clickable to display to the user what the world is saying by showing her the top 20 search results on the topic of that tweet. This feature is included to observe the clicking activity of the user to assess the the user's interest in reading more about the world or to see why the world lies where it does along the bar (Figure 1). This visualization enables social comparison: it gives the user the opportunity to compare the political slant of what she reads and what other Twitter users read on a political topic.

*Feed Bias:* In addition to the political bias of each tweet and the Twitter world, we calculated the overall bias of a user's feed by computing the average political bias of all the political tweets in her feed. We also calculated the average bias of what other Twitter users read about the political topics discussed in the user's feed. These two biases are displayed to the user on an overall bias bar at the top dock of the screen. These biases reflect to the user the overall biases of what they see on their feeds compared to what the world is saying which could highlight the possible filter bubble they are in (Figure2).

### Experiment Design

We conducted an in-lab interview with nine Twitter users. We asked participants to explore their feed via the tool and engaged them in an open-ended discussion to understand their opinions and attitudes towards the visualizations the tool provided.

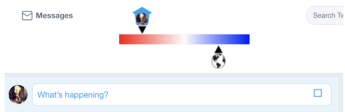

**Figure 2:** The overall Feed's bias versus the overall Twitter world's bias

*Pre-interview*
Before showing the tool to the participant, we asked participants to fill a short survey asking about some demographic data and political activity. We then asked participants to to estimate their overall feed bias and the overall Twitter World's bias on a 7 point Likert scale, along with their satisfaction with their overall feed bias and if they wanted to change it. These questions measured the participant's awareness of what they read on their feed in comparison of what the world (or wider community) is saying. We also asked participants how much they read news from the opposing party, how satisfied they were with their existing tweet consumption and if they wanted to change it. We repeated these questions at the end of the interview to measure the intervention effect of the tool.

*Tweets' Bias Evaluation*
Next, we walked participants through the tool and explained its features to them. We asked them to explore their feed via the tool, and engaged them in an open-ended discussion about each political tweet in their feed. We particularly asked participants to discuss their opinion about the political bias of each tweet versus the Twitter world bias on that topic. Each participant evaluated 13 political posts on average.

*Feed's Bias Evaluation*
Finally, we asked participants to check their overall feed bias and compare it to the world (Figure2), rate their satisfaction with their overall feed bias, and mention if they wanted to change it. Comparing the actual overall biases with the ones estimated in the beginning of the interview reflect the participant's awareness of their filter bubble.

*Participants*
We recruited nine participants in the Bay Area along different education levels (33.3% Bachelor's degree, 22.2%

Master's degree, 11.1% Associate's degree, 33.3% some college or no degree). 55.6% of participants were from 18-24 and the rest were from 25 - 34. 44.4% were female. Participants belonged to three ethnic groups (11.1% African American, 55.6% Asian, 33.3% white). The party affiliation distribution of participants was 22.2% weak republican, 11.1% weak democrat, 44.4% moderate democrat and 22.2% strong democrat. Participants' income were distributed widely from less than $10,000 to $150,000.

## Findings and Discussion

### Comments on the Tool
Visually introducing the idea of social comparison to participants interested many of them. They showed positive reactions when they explored the tool. Some giggled out of surprise or dropped comments such as "*This is very interesting!*" (P8), "*Oh! I didn't know this is a thing. That is so cool*" (P2) and "*Oh! Wow*" (P4). Such comments reflected how this visual comparison with the Twitter world could be mentally provocative.

### Exposure to Opposing Content
During the pre-interview, eight participants highlighted that it is necessary to read news from the opposing party "*because if you just take info from one side then your bias comes at one side. You need to understand how the other person thinks to be able to convince yourself*" (P7). They reflected high awareness of the filter bubble phenomenon: "*you got people who are living in their bubble, Californians living in their California bubble, Kentucky people living in their Kentucky bubble. You have a bubble of geography, you have a bubble of politics*" (P7). Therefore, "*you should know what is going on and you should not be caught in this bubble*" (P4). They argued that reading from the opposing political party "*helps to see why you could be wrong. If you still believe you are right, it gives you a good way to talk to and understand people on the other side and have better conversations with them*" (P8).

One participant also reflected on the impact of filtering algorithms in reinforcing the filter bubble: "*Social media has started creating profiles on you and started pushing certain content along your way that you would want to see, so I think sometimes when I go into an anonymous mode, I definitely see more content that is leaning more liberal [which is the opposing party]*" (P7).

### The Desire to Change
While almost all the participants were aware of the importance of reading opposing content during the pre-interview, most of them (n=7) still stated that they did not want/care to change their reading rate of opposing content "*just because of the current political climate. Every time I read something from the other party, it is usually very strong and traditional. It spares a lot of emotions and spares up a lot of controversy in things I don't necessarily agree with. I would rather not ruin my day by reading another article*" (P4). They particularly argued that reading content from the opposite political party bothered them: "*I don't really feel like I can read more than a certain amount. I have to read the opposite party in small bits*" (P7).

This corroborates previous findings that showed while people are usually aware of the filter bubble, they are usually challenge-averse [12]; i.e., they do not want to change their reading behavior. However, there were participants who still wanted to change their reading rate of the opposing content if they were provided the right content: "*I would rather have something from the other party suggested to me rather than following more people from the other party*" (P9). This shows that while people do not want their feeds to be loaded with opposing content, some who are diversity-seeking [12] might still want to know about the other party's news. This highlights the possible effectiveness of social comparison as a window to explore more opinions.

***The Impact of Social Comparison***
Even though most of participants stated that they do not want/care to change their reading behavior before walking through the tool, we found that more than half of the participants (n=5) desired to change their reading rate of opposing content after being exposed to what other Twitter users read on various political topics. They stated that they were "*okay with watching more things from the opposite party*" (P1) after comparing what they read and what other users read. Some also highlighted that they "*need to diversify more on the opposite side*" (P7) when they noticed that what they were currently reading was only extreme content from the opposite political party. Therefore, they stated that they "*would like to get more civilized other party tweets*" (P8). This shows that social comparison can raise a user's awareness of what content they consume and help them to re-evaluate their reading behavior.

*Clicking Activity*
We found that more than half of the participants (n=6) clicked or showed interest in clicking the world icon in at least 60% of their political tweets. This interest was mainly affected by the position of the Twitter world bias compared to the tweet's bias: participants were mostly satisfied when what they read and what others read had the same political slant. Participants highlighted that they felt "*happy*" (P6), and "*interested*" (P5, P9) when the world was on their side, and they felt "*alarmed, concerned*" (P1), "*disheartened, disgusted, angry, pissed off*" (P4), and "*shocked*" (P8) when the world was on the opposite side of their tweets' political affiliation. Seeing the world on the same side as their political affiliation, participants "*felt better*" (P4), and more positive and assured "*because it matches their ideas and what they agree with*" (P5).

Such emotions provoked participants to click the world icon when the world was on the side of the opposing political party: "*If the world is on the opposite side, I would be very angry. I will definitely click the world! I wouldn't click it otherwise*" (P4). Participants also stated that if the world is on the other side from the tweet on the bar, they "*would read it out of curiosity*" (P3) to understand "*what the other side is saying*" (P7). Most of the participants asserted that they did not want to click on the world icon if the world and the tweet's bias were aligned because they already "*have enough context*" (P3) on that topic.

Participants clicked the world icon *and* carefully explored what the world was saying. One participant (P7), who stated that he had a certain capacity for reading from the opposite party, read even more world-representng articles . His curiosity and other triggered emotions resulting from the position of the world icon drove them to compare the tweets of the world with the position of the world on the bar and with the tweet itself. This showed that the presence of the world measure provoked participants to read more due to social comparison and the fact that the world content is "*the most viral and people are sharing it*" (P5).

***Overall Feed Bias versus the World Bias***
Similar to comparing a tweet's bias to the Twitter world bias on a topic, we found that the more a participant's overall feed bias was close to the Twitter world bias, the more they were satisfied. Eight participants expressed their satisfaction when the Twitter world was on their political side and their dissatisfaction when the world was on the opposite side: "*I would feel bitter about the world in the opposite side, glad that the world is on my side*" (P4). They even used the world bias as their reference: "*I want the world to be very extreme and me chasing it on the bar*" (P6). There-fore, if the world was not on the same political side that participants belonged to, they stated that "*some actions must be taken*" (P6), for example by tweeting their viewpoints more, to help balance the world. This highlights the effectiveness of social comparison on users' satisfaction and even reading behavior.

However, is it reasonable to always consider other users (e.g., the Twitter world here) as a reference? What if the world itself is biased towards a wrong political opinion on a topic (e.g., society preventing black people from the rights white people have had for years)? Or what if being close to the world's political slant means that a user is in a filter bubble? These concerns were reflected in a participant's comments about being on the world's side: "*If the world is on the same side as me, I would have probably been living in a bubble*" (P8). This shows that while social comparison can be a powerful driver for users to reflect on their reading behavior, further research is needed to better understand how it might affect users' reading behavior adversely, and what design solutions can mitigate such issues.

## Limitations

This was an exploratory study with only a few participants to conduct an initial evaluation of the tool we built and the impacts of social comparison on users' political reading behavior. We hope that the results of this exploratory study can motivate further research with a larger set of users and more features (such as changes in users' reading behavior in the long term).

The method we used to calculate the political bias of a tweet can be also improved. During the study, some of the tweets' political slants seemed to be a bit miscalculated for some participants; it did not make complete sense to them. Implementing a model with accuracy in estimating the tweet bias would mitigate this confusion.

## Conclusions & Future Work

In this case study, we showed that social comparison can have a great impact on users' political reading behavior by motivating them to read diverse political opinions. While previous work used self-comparison as a nudging practice to break users' political filter bubble, our study shows that social comparison can be a much stronger motive for users, specifically because people tend to compare themselves with others to define their own self.

This work, however, was a preliminary step in evaluating the impacts of social comparison in breaking the filter bubble. We hope that by converting our tool to a plugin that every Twitter user can choose to use, we can study users' activity to understand the effects of social comparison on users' reading behavior longitudinally.